\documentclass{iopart}
\usepackage{bm}
\usepackage{graphicx}

\newcommand{\sump}[1]{{\sum_{#1}}\raise4pt\hbox{$'$}}

\begin{document}

\title[Sherrington-Kirkpatrick model near $T=T_c$]%
      {Sherrington-Kirkpatrick model near $T=T_c$: 
expanding around the Replica Symmetric Solution}

\author{A Crisanti$^1$ and C De Dominicis$^2$}

\address{$^1$ Dipartimento di Fisica, Universit\`a di Roma 
              {\em La Sapienza} and  SMC, 
              P.le Aldo Moro 2, I-00185 Roma, Italy.
        }
\address{$^2$ {\em Institut de Physique Th\'eorique}, CEA -
              Saclay - Orme des Merisiers, 91191 Gif sur Yvette, 
              France
        }
\eads{\mailto{andrea.crisanti@phys.uniroma1.it}, \mailto{cirano.de-dominicis@cea.fr}
     }

\begin{abstract}
An expansion for the free energy functional of the Sherrington-Kirkpatrick 
(SK) model, around the Replica Symmetric (RS) SK solution 
$Q^{({\rm RS})}_{ab} = \delta_{ab} + q(1-\delta_{ab})$
is investigated.
In particular, when the expansion is truncated to fourth order in.
$Q_{ab} - Q^{({\rm RS})}_{ab}$. 
The Full Replica Symmetry Broken (FRSB) solution is explicitly found
but it turns out to exist only in the range of temperature 
$0.549\ldots\leq T\leq T_c=1$, not including $T=0$. On the other hand an 
expansion around the paramagnetic solution
$Q^{({\rm PM})}_{ab} = \delta_{ab}$ up to fourth order yields a FRSB solution
that exists in a limited temperature range 
$0.915\ldots\leq T \leq T_c=1$.
\end{abstract}
\submitto{\JPA}

\date{V 3.1.3 2010/01/18 08:27:39 AC} 
\maketitle

\section{Introduction:}
The Sherrington-Kirkpatrick (SK) model is defined by the Hamiltonian 
\cite{SheKir78}:
\begin{equation}
{\cal H} = -\frac{1}{2}\sum_{i\not=j}^{1,N}\,J_{ij}\,\sigma_i\sigma_j
\end{equation}
where the $\sigma_i$ are $\pm 1$ Ising spins and the couplings $J_{ij}$ 
are independent Gaussian random variables of zero mean and variance 
equal to $1/N$. 

The thermodynamic properties of the model are described by the
free energy (density) $f$ averaged over the quenched disorder. 
To overcame the difficulties of averaging a logarithm, 
the average over the disorder is
computed using the so called {\sl replica trick}:
\begin{equation}
 -\beta N\overline{f} = \lim_{n\to 0}\frac{\overline{Z^n}-1}{n}
\end{equation}
where $\beta = 1/T$ is the inverse temperature and, 
as usual, $\overline{(\cdots)}$ denotes the average over the disorder.
For $n$ integer $Z^n$ is the partition functions of $n$ identical, 
non interacting, replicas of the system. The average over disorder
couples the different replicas.
Performing this average, and 
introducing the auxiliary symmetric replica overlap matrix 
$Q_{ab} = \frac{1}{N}\sum_i\sigma_{ia}\sigma_{ib}$, with $a\not=b$,
the disorder averaged replicated partition functions can be written as
\cite{SheKir78}:
\begin{eqnarray}
\overline{Z^n}= \int \prod_{a<b} \sqrt{\frac{N\beta^2}{2\pi}}\,\rmd Q_{ab}
             \, \rme^{N {\cal L}[Q]}
\label{eq:lag}
\end{eqnarray}
with the effective Lagrangian (density):
\begin{eqnarray}
{\cal L}[Q]&=& -\frac{\beta^2}{4} \sum_{ab} Q_{ab}^2 
                       + \Omega[Q]
                       - n\frac{\beta^2}{4}
\label{eq:l} \\
\Omega[Q] &=&\ln \Tr_{\sigma_a} \exp\bigg(\frac{\beta^2}{2}\sum_{ab} 
             Q_{ab}\, \sigma_a\sigma_b\bigg)
\label{eq:ome}
\end{eqnarray}
The last term in (\ref{eq:l}) follows from the definition
$Q_{aa} = 1$. 
The normalization factor in (\ref{eq:lag}) gives a sub-leading 
contributions for $N\to\infty$ and is omitted in the following.

In the thermodynamic limit, $N\to\infty$, the value of the integral in 
(\ref{eq:lag}) is given by the stationary point value, and the replica 
free energy density reads:
\begin{equation}
\label{eq:freen}
  -n\beta\,f = {\cal L}[Q]
\end{equation}
with $Q_{ab}$ evaluated from the stationary condition
\begin{equation}
\frac{\partial}{\partial\,Q_{ab}}\,{\cal L}[Q] = 0, \quad
a < b
\end{equation}
that is from the self-consistent equation
\begin{equation}
  Q_{ab} 
         = \frac{\Tr_{\bm\sigma} \sigma_a\sigma_b\,
      \exp\left(\frac{\beta^2}{2}\sum_{ab} Q_{ab}\, \sigma_a\sigma_b\right)}
         { \Tr_{\bm\sigma}
      \exp\left(\frac{\beta^2}{2}\sum_{ab} Q_{ab}\, \sigma_a\sigma_b\right)}
= \langle\sigma_a\sigma_b\rangle, \qquad
a \not= b.
\end{equation}
To solve the self-consistent stationary point equation we have to specify the 
structure of the matrix $Q_{ab}$. This is not straightforward since the 
symmetry of the replicated partition function under replica permutation 
is broken in the low temperature phase. 
The Replica Symmetric (RS) {\it Ansatz} $Q_{ab} = \delta_{ab} + q\,(1-\delta_{ab})$
of Sherrington-Kirkpatrick \cite{SheKir78}, 
that assumes the same overlap for any pair of 
replicas, indeed yields an unphysical negative entropy at zero 
temperature.
Following the parameterization introduced by Parisi \cite{Parisi79, Parisi80}, 
the overlap matrix
$Q_{ab}$ for $R$ breaking in the replica permutation symmetry is 
divided into successive boxes of decreasing size $p_r$, 
with $p_0 = n$ and $p_{R+1}=1$, along the diagonal, and
the elements $Q_{ab}$ of the overlap matrix are assigned so that
\begin{equation}
Q_{ab} \equiv q_{a\cap b=r} = Q_r, \qquad r = 0,\cdots, R+1
\end{equation}
with $1=Q_{R+1}\geq Q_R \geq\cdots \geq Q_1 > Q_0$.
The notation 
$a\cap b=r$ means that $a$ and $b$ belong to the 
same box of size $p_r$ but to two {\it distinct} boxes of size $p_{r+1} < p_r$.
The case $R=0$ gives back the RS solution, while the opposit limit
$R\to\infty$ describes a state with an infinite, continuum, number of possible
spontaneous breaking of the replica permutation symmetry.
It turns out that a physical solution is obtained only in the latter case.
Using this structure for $Q_{ab}$, Parisi and 
others \cite{Parisi79,Parisi80,Duplantier81} have shown 
how to obtain solutions with $R$ steps of replica symmetry breaking (RSB) and 
in particular with $R\to\infty$ (FRSB), 
and how to construct equations satisfied by 
$Q(x)$, the continuous limit of the order parameter $Q_{ab}$
for $R\to \infty$ \cite{Parisi83}. 
These equations can be solved in the full low temperature
phase \cite{CriRiz02,CriRizTem03,Pankov06,OppShe05,OppSchShe07,SchOpp08}.
However working directly with $Q(x)$ makes it difficult to keep track,
for instance, of the Hessian, and hence of the stability of the
solution, since the matrix structure of the overlap matrix
$Q_{ab}$ is lost in the continuous limit.\footnote{Paradoxically, it is this 
continuous limit $R\to\infty$, that imposes the existence of zero modes (at 
the bottom of the replicon bands). Indeed, this limit is necessary
to transform the replica permutation invariance into a (broken) 
{\sl continuous} group thus generating Goldstone zero modes.} 
The study of the Hessian of the fluctuations around the RSB solution 
with an arbitrary $R$ from the Lagrangean (\ref{eq:l})-(\ref{eq:ome})
is a very hard task.
As a result stability analysis has been mostly investigated near the
critical temperature and with the help of a simplified 
model \cite{BraMoo78,DeDomKon83}, the so called {\sl Truncated Model}
\cite{Parisi79,BraMoo79},
that similarly to the Landau Lagrangian retains only the main mathematical
structure of the expansion of the replicated free energy 
in powers of $Q_{ab}$ 
near $T_c$, where $|Q_{ab}| \ll 1$.

In the present work, we take a different viewpoint and consider
the expansion of the Lagrangean (\ref{eq:l})-(\ref{eq:ome})
around the Replica Symmetric {\sl ansatz} of Sherrington and 
Kirkpatrick. The main motivation for such an expansion is to 
obtain a simpler Lagrangean which, while retaining the replica
symmetry breaking properties of the original model, 
is {\it a priori} valid in the whole low temperature phase.
Anticipating our conclusions, we find that the model obtained by truncating the
expansion to the fourth order, the minimum order required to have a 
FRSB solution, while improving the results obtained from the 
expansions near $T_c$ is valid in a temperature range which does not reach
zero temperature. 

The outline of the paper is as follows: in Section \ref{sec:expa} we construct 
the approximation of $\Omega[Q]$ obtained expanding it around the Replica 
Symmetric SK solution $Q_{ab}^{(RS)} = q$ ($a\not=b$) up to fourth order 
in $Q_{ab} - q$.
The stationarity equation and its solutions are discussed in 
Section \ref{sec:steq}. 
The Truncated Model was obtained considering the main features 
of the mathematical structure of the
expansion of $\Omega[Q]$ around the paramagnetic solution $Q_{ab}^{(PM)}=0$ 
($a\not=b$) to
fourth order in $Q_{ab}$. The parameters entering in the model are, however,
usually arbitrary and so it is difficult to make contact with the original
SK model. By using the results of Section \ref{sec:expa}
we can determine the coefficients of the expansion and study 
the properties of the solution. This is done in Section \ref{sec:expa0}.
Discussion and conclusions are deferred to Section \ref{sec:conc}.

\section{Expansion of the free energy functional around the SK solution:}
\label{sec:expa}
To expand the functional $\Omega[Q]$ around the SK solution 
$Q_{ab} = q$ for $a\not=b$, we consider an overlap matrix $Q_{ab}$ of the form
\begin{equation}
\label{eq:Q}
Q_{ab} = \delta_{ab} + q\,(1-\delta_{ab}) + q_{ab}
\end{equation}
where $q$ is given by the SK Replica Symmetric solution (see below) and
$q_{ab}$ the deviation from the Replica Symmetric solution. 
Inserting this form of $Q_{ab}$ into the free energy functional 
(\ref{eq:freen}) yields:
\begin{eqnarray}
\label{eq:freenq}
-n\beta f &=& n\frac{\beta^2}{4}\,q^2 - n \frac{\beta^2}{2}\,q 
            -\frac{\beta^2}{2} q \sum_{ab}q_{ab} 
            -\frac{\beta^2}{4}\sum_{ab}q_{ab}^2 
\nonumber\\
&\phantom{=}&
                  + \ln\Tr_{\bm \sigma}\,
   \exp\left[\frac{\beta^2}{2}q\left(\sum_{ab}\sigma_a\right)^2 + 
             \frac{\beta^2}{2}\sum_{ab}^{1,n}q_{ab}\,\sigma_a\sigma_b\right]
             + \Or(n^2)
\end{eqnarray}
Setting $q_{ab}=0$ the above expression leads to the Sherrington-Kirkpatrick
free energy
\begin{equation}
\label{eq:freenSK}
-\beta f_{\rm SK} = \frac{\beta^2}{4}q^2 - \frac{\beta^2}{2}q 
           + \overline{\ln\cosh(\beta z)} + \ln 2
             + \Or(n)
\end{equation}
where the overbar denotes the average over the Gaussian variable $z$:
\begin{equation}
\overline{g(z)} = \int_{-\infty}^{+\infty}\frac{\rmd z}{\sqrt{2\pi q}}
           \rme^{-z^2/2q}\,g(z).
\end{equation}
Stationarity of $f_{\rm SK}$ with respect to $q$ leads to
SK Replica Symmetric solution:
\begin{equation}
\label{eq:qSK}
  q = \overline{\theta^2}, \qquad \theta \equiv \tanh(\beta z).
\end{equation}

For $q_{ab}\not=0$ the free energy functional $f$ can be written,
expanding the last term in (\ref{eq:freenq}) in powers of $q_{ab}$, 
as:
\begin{eqnarray}
\label{eq:freenc}
\fl
-n\beta f = -n\beta f_{\rm SK} 
            -\frac{\beta^2}{2} q \sum_{ab}q_{ab} 
            -\frac{\beta^2}{4}\sum_{ab}q_{ab}^2 
\nonumber\\
 + \sum_{k\geq 1} \frac{1}{k!}\left(\frac{\beta^2}{2}\right)^k 
    \left\langle\left(\sum_{ab}q_{ab}\,\sigma_a\sigma_b\right)^k\right\rangle_c
 \end{eqnarray}
where the subscript ``$c$'' indicates that only connected contributions, i.e., 
only those terms that cannot be written as the product of two or more 
independent sums, must be considered.
The angular brackets denote the average
\begin{equation}
  \langle g({\bm \sigma})\rangle = 
        \overline{\prod_{a=1}^n e^{\beta z\sigma_a}g({\bm \sigma})}
        + \Or(n).
\end{equation}
Since $\sigma_a^2 = 1$, 
the last term in (\ref{eq:freenc}) 
contains only averages of products of spins with different replica index. 
These are easily evaluated yielding
\begin{eqnarray}
\langle\sigma_{a_1}\cdots\sigma_{a_h}\rangle &= 
        \overline{\prod_{a=1}^n e^{\beta z\sigma_a} \prod_{l=1}^h\sigma_l
                 }
\nonumber\\
  &=         \overline{
              \left[2\cosh(\beta z)\right]^{n-h}
              \left[2\sinh(\beta z)\right]^h
                 } + \Or(n)
\nonumber\\
  &= \overline{\theta^h} + \Or(n), \qquad\qquad\qquad a_1\not=\cdots\not= a_h.
\end{eqnarray}
Form the study of the truncated model it is known that terms of order
$\Or(q_{ab}^4)$ must be included into the free energy to break the replica 
symmetry. Thus in the following we shall consider the first four terms of 
the expansion.

\subsubsection{Term $\Or(q_{ab})$:}
The term of order $\Or(q_{ab})$ is
\begin{equation}
\left\langle\sum_{ab}q_{ab}\,\sigma_a\sigma_b\right\rangle = 
     \sum_{ab}q_{ab}\langle\sigma_a\sigma_b\rangle 
    = \overline{\theta^2}\sum_{ab}q_{ab}
\end{equation}
The choice $q=\overline{\theta^2}$, 
see (\ref{eq:qSK}), cancels the linear term in the expansion 
(\ref{eq:freenc}) and removes the tad-poles.

\subsubsection{Terms $\Or(q_{ab}^2)$:}
The term of order $\Or(q_{ab}^2)$ reads
\begin{equation}
\label{eq:o2}
\left\langle\left(\sum_{ab}q_{ab}\,\sigma_a\sigma_b\right)^2\right\rangle = 
    \sum_{ab\atop cd} q_{ab}\,q_{cd}\,
          \langle\sigma_a\sigma_b\sigma_c\sigma_d\rangle
\end{equation}
To evaluate this term we have to find all different possible ways
of equating the $ab$ indexes to $cd$ indexes with the constraint, 
imposed by $q_{aa}=0$, that $a\not= b$ and $c\not= d$. 
There are three possible cases: all indexes different, 
a pair of equal indexes, and two pairs of equal indexes. 
By noticing that the spin product averages depend
only on the number of different indexes, and not on the value of the indexes, 
and that the matrix $q_{ab}$ is symmetric, these yield
\begin{eqnarray}
\fl
\label{eq:q2rest}
\left\langle\left(\sum_{ab}q_{ab}\,\sigma_a\sigma_b\right)^2\right\rangle =
      \overline{\theta^4}\,{\sum_{abcd}}' q_{ab}\,q_{cd}
  + 4\,\overline{\theta^2}\,{\sum_{abc}}' q_{ac}\,q_{cb}
  + 2 {\sum_{ab}}' q_{ab}^2,
\end{eqnarray}
since there are $4$ possible ways of equating one index in 
$ab$ with one index in $cd$ and  $2$ was of equating the pair of indexes $ab$ 
to the pair $cd$. All sums are restricted to different indexes, this is 
denoted by the prime ``$'$'' over the sum sign. 
Transforming the restricted sums into unrestricted ones, i.e.,  
sums over free index, one finally ends up with:
\begin{eqnarray}
\fl
\label{eq:q2unresta}
\left\langle\left(\sum_{ab}q_{ab}\,\sigma_a\sigma_b\right)^2\right\rangle =
        \overline{\theta^4}\ \sum_{abcd}q_{ab}\,q_{cd}
   &+ 4\,\overline{\theta^2(1-\theta^2)}\ \sum_{abc}q_{ac}\,q_{cb}
\nonumber\\
   &+ 2\,\overline{(1-\theta^2)^2}\ \sum_{ab}q_{ab}^2
\end{eqnarray}
This equation has a simple diagrammatic expression. 
Indeed denoting $q_{ab}$ by a line and the vertex where
two (or more) indexes are equal by a ``dot'', the above equation can be 
written as
\begin{eqnarray}
\fl
\label{eq:q2unrestd}
\left\langle\left(\sum_{ab}q_{ab}\,\sigma_a\sigma_b\right)^2\right\rangle =
  \overline{\theta^4}\ \includegraphics[bb= 106 674 125 695]{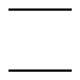}
 &+ 4\,\overline{\theta^2(1-\theta^2)}\ %
\includegraphics[bb= 105 675 126 694]{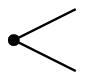}
\nonumber\\
 &+ 2\,\overline{(1-\theta^2)^2}\ %
\includegraphics[bb= 106 683 147 702]{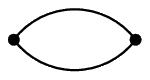}
\end{eqnarray}
More details can be found in \ref{app:O2det}.
From this form we easily see that the first term is a disconnected contribution
and hence it does not appears in the free energy (\ref{eq:freenc}), 
therefore to order $\Or(q_{ab}^2)$ the free energy reads
\begin{equation}
\label{eq:freen2c}
\fl
-n\beta f = -n\beta f_{\rm SK} 
    + \frac{\beta^4}{4}M\,\sum_{abc}q_{ac}q_{cb} 
    + \frac{\beta^4}{4} N\,\sum_{ab}q_{ab}^2 
+ \Or(n^2,q_{ab}^3)
 \end{equation}
where
\begin{equation}
M = 2\,\overline{\theta^2(1-\theta^2)}, \quad
N = \overline{(1-\theta^2)^2} - T^2.
\end{equation}
Notice that the coefficient $N$ is (minus) the Replicon eigenvalue of the 
Replica Symmetric solution \cite{deAlmTho78}. The $q_{ab}=0$ solution is
hence unstable below $T=1$.

\subsubsection{Terms $\Or(q_{ab}^3)$ and $\Or(q_{ab}^4)$:}
These are evaluated as done for the $\Or(q_{ab}^2)$ by computing all connected 
contributions that follows from the expansion of the $k=3$ and $k=4$ terms in 
(\ref{eq:freenc}). By using a self-explanatory diagrammatic representation 
these are given by:
\begin{equation}
\fl
\label{eq:freeq3d}
\left\langle\left(\sum_{ab}q_{ab}\,\sigma_a\sigma_b\right)^3\right\rangle_c =
     P\ \includegraphics[bb= 97 679 125 693]{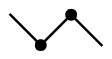}
  +  Q\ \includegraphics[bb= 97 679 116 691]{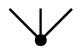}
  +  R\ \includegraphics[bb= 104 684 143 697]{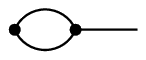}
  + J\ \includegraphics[bb= 104 684 127 697]{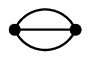}
  + K\ \includegraphics[bb= 104 679 127 699]{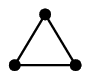}
\end{equation}
where 
\begin{equation}
\fl
    P =  24\,\overline{\theta^2(1-\theta^2)^2}  ,\quad
    Q = -16\,\overline{\theta^4(1-\theta^2)}    ,\quad
    R = -48\,\overline{\theta^2(1-\theta^2)^2}  ,\quad
\end{equation}
\begin{equation}
\fl
    J =  16\,\overline{\theta^2(1-\theta^2)^2}  ,\quad
    K =   8\,\overline{(1-\theta^2)^3}
\end{equation}
and
\begin{eqnarray}
\fl
\label{eq:freeq4d}
\left\langle\left(\sum_{ab}q_{ab}\,\sigma_a\sigma_b\right)^4\right\rangle_c =
 &-  A\ \includegraphics[bb= 94 679 119 691]{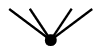}
  +  B\ \includegraphics[bb= 88 684 143 697]{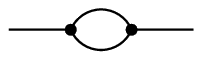}
  -  B\ \includegraphics[bb= 97 679 125 693]{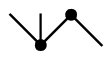}
\nonumber\\
 &+  C\ \includegraphics[bb= 95 682 122 705]{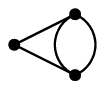}
  -  C\ \includegraphics[bb= 104 679 143 699]{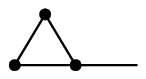}
  +  4D\ \includegraphics[bb= 104 683 143 694]{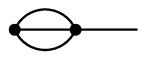}
\nonumber\\
 &-  3D\ \includegraphics[bb= 104 681 143 700]{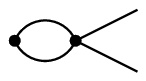}
  +  E\ \includegraphics[bb= 97 678 134 692]{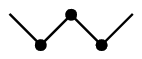}
  -  2E\ \includegraphics[bb= 104 679 143 697]{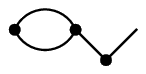}
\nonumber\\
 &+  F\ \includegraphics[bb= 104 682 127 705]{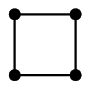}
  +  G\ \includegraphics[bb= 104 682 127 709]{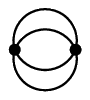}
  -  H\ \includegraphics[bb= 104 683 144 694]{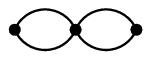}
\end{eqnarray}
with
\begin{equation}
\fl
A =  32\,\overline{\theta^4(1-3\theta^2)(1-\theta^2)}, \quad
B = 384\,\overline{\theta^4(1-\theta^2)^2}, \quad
C = 384\,\overline{\theta^2(1-\theta^2)^3},
\end{equation}
\begin{equation}
\fl
D =  64\,\overline{\theta^2(1-3\theta^2)(1-\theta^2)^2}, \quad
E = 192\,\overline{\theta^2(1-\theta^2)^2}, \quad
F =  48\,\overline{(1-\theta^2)^4},
\end{equation}
\begin{equation}
\fl
G =  32\,\overline{(1-3\theta^2)^2(1-\theta^2)^2}, \quad
H =  96\,\overline{(1-3\theta^2)(1-\theta^2)^3}.
\end{equation}
Collecting all contributions up to order $\Or(q_{ab}^4)$, the replica free 
energy functional reads:
\begin{eqnarray}
\fl
\label{eq:freen4c}
-n\beta f = -n\beta f_{\rm SK} 
  +\frac{1}{4T^4}
            \left[M\,\sum_{abc}q_{ac}q_{cb} + N\, \sum_{ab}q_{ab}^2\right]
   +\frac{1}{6(2T^2)^3}
            \left[
    P\,\sum_{abcd}q_{ac}q_{cd}q_{db} 
    \right.
\nonumber\\
  \left.
  + Q\,\sum_{abcd} q_{ad}q_{bd}q_{cd}
  + R\,\sum_{abc} q_{ac}^2q_{cb} 
  + J\,\sum_{ab}q_{ab}^3 
  + K\, \sum_{abc} q_{ac}q_{cb}q_{ba}
            \right]
\nonumber\\
  +\frac{1}{24(2T^2)^4}
            \left[
   - A\,\sum_{abcde}q_{ae}q_{be}q_{ce}q_{de}
   + B\,\sum_{abcd} q_{ac}q_{cd}^2q_{db}
   \right.
\\
   - B\,\sum_{abcde}q_{ac}q_{dc}q_{ce}q_{eb}
   + C\,\sum_{abc} q_{ac}q_{cb}^2q_{ba}
   - C\,\sum_{abcd}q_{ac}q_{ad}q_{dc}q_{cb}
\nonumber\\
   +4D\,\sum_{abc}q_{ac}^3q_{cb}
   -3D\,\sum_{abcd}q_{ab}^2q_{bc}q_{bd}
   +E\,\sum_{abcde}q_{ab}q_{bc}q_{cd}q_{de}
\nonumber\\
   \left.
   -2E\,\sum_{abcd}q_{ab}q_{bc}q_{cd}^2
   +F\,\sum_{abcd}q_{ab}q_{bc}q_{cd}q_{da}
   + G\,\sum_{ab}q_{ab}^4 
   - H\,\sum_{abc}q_{ac}^2q_{cb}^2
   \right]
\nonumber\\
+ \Or(n^2,q_{ab}^5)
\nonumber
\end{eqnarray}

\section{Stationarity equation:}
\label{sec:steq}
The equation for $q_{ab}$ follows from the stationarity condition 
$(\partial/\partial q_{ab}) f = 0$ applied to the replica free energy 
functional
(\ref{eq:freen4c}).  In the limit $R\to\infty$ this yields 
\begin{eqnarray}
\fl
\label{eq:speq}
\frac{1}{2T^4}\left[ M S_1 + Nq(x) \right]
+\frac{1}{6(2T^2)^3}
 \Biggl[
        3(P+Q)S_1^2  + R \Bigl(S_2 + 2 S_1 q(x)\Bigr) + 3 J q(x)^2
\nonumber\\
   + 6 K\left(\int_{0}^{x}\rmd y\,\dot{q}(y)\,\widehat{q}(y) + S_1 q(0)\right)
 \Biggr]
+\frac{1}{24(2T^2)^4}
 \Biggl[
    -4AS_1^3
\nonumber\\
  + B\Bigl(2S_1S_2 - 4 S_1^3 + 2 S_1^2 q(x)\Bigr)
   + C\,\Delta(x)
\nonumber\\
   + D\Bigl(4S_3 - 6S_1S_2 + 12 S_1 q(x)^2 - 6 S_1^2q(x)\Bigr)
\nonumber\\
   + E\Bigl(4S_1^3 - 4 S_1 S_2 - 4 S_1^2 q(x)\Bigr)
\nonumber\\
   + 12F\left(\int_{0}^{x}\rmd y\, \dot{q}(y)\,\widehat{q}(y)^2 
        + S_1^2 q(0)\right)
\nonumber\\
   + 4 G\, q(x)^3 - 4HS_2\, q(x)
 \Biggr] 
= 0,
\qquad 0\leq x\leq x_c,
\end{eqnarray}
where 
\begin{eqnarray}
\fl
\Delta(x)= 2\left[\int_{0}^x\rmd y\left(\frac{\rmd}{\rmd y}q(y)^2\,
                           \widehat{q}(y) +
                         \widehat{q^2}(y)\,\dot{q}(y)\right) 
                  + S_1\,q(0)^2 + S_2\, q(0)
\right]
\nonumber\\
+\Bigl(4q(x)-6S_1\Bigr)\left[\int_0^x\rmd y\,\dot{q}(y)\,\widehat{q}(y) 
                             + S_1\,q(0)
                           \right]
\nonumber\\
  -3 \int_0^x\rmd y\, q(y)\,\widehat{q}(y)^2 
   - 3 S_1^2\,q(0) + q(x)^3
\end{eqnarray}
and
\begin{equation}
\label{eq:sn}
S_n = -\int_{0}^1 \rmd x\, q(x)^n
    = -\int_{0}^{x_c} \rmd x\, q(x)^n - (1-x_c)\,q(x_c)^n
\end{equation}
The ``dot'' indicates the derivative, $\dot{q}(x) = (d/dx)q(x)$, while the
``hat'' the Replica Fourier Transform (RFT), that for $R\to\infty$ reads
\cite{DeDomCarTem97}:\footnote{The RFT was first introduced, 
directly in the continuum limit ($R\to\infty$) by Mezard and Parisi
\cite{MezPar91}}

\begin{equation}
\widehat{q}(x) = \int_{x}^{x_c}\rmd y\,y\frac{\rmd}{\rmd y}q(y)-q(x_c), 
\qquad\mbox{\rm RFT}
\end{equation}
\begin{equation}
q(x) = -\int_{0}^{x} \rmd y\,\frac{1}{y}\frac{\rmd}{\rmd y}\widehat{q}(y)  + q(0)
\qquad\mbox{\rm inverse RFT}
\end{equation}
where $q(0) = q(x=0)$, and we have neglected the surface term at $x=1$
since $q(x=1) = q_{aa} = 0$.

\subsection{Solution of the Stationarity equation}
\label{sec:solu}
The complicate integro-differential stationarity equation (\ref{eq:speq}) 
can be solved reducing it to an ordinary differential equations 
using 
the differential operator $\widehat{\bm O}= (1/\dot{q}(x))(\rmd/\rmd x)$ 
to eliminate integrals.
Application of $\widehat{\bm O}$ to (\ref{eq:speq})
leads to
\begin{eqnarray}
\label{eq:speqd1}
\fl
\frac{N}{2T^2}
+ \frac{1}{3(2T^2)^3}\biggl[RS_1 + 3Jq(x) + 3K \widehat{q}(x)\biggr]
+ \frac{1}{12(2T^2)^4}\biggl[BS_1^2 
\nonumber\\
+ C\biggl(2\int_{0}^x\rmd y\,\dot{q}(y)\,\widehat{q}(y) + \widehat{q^2}(x) 
         +4q(x)\widehat{q}(x) - 3 S_1\,\widehat{q}(x) 
\nonumber\\
+ 2S_1\,q(0)\biggr)
            +D\bigl(12 S_1 q(x) - 3 S_1^2\bigr) 
    -2ES_1^2 + 6F\widehat{q}(x)^2 
\nonumber\\
+ 6Gq(x)^2 - 2HS_2\biggr]
= 0.
\end{eqnarray}
The equation is not yet simple enough to be solved. A second 
application of $\widehat{\bm O}$, and a rearrangement of terms, yields
\begin{equation}
\label{eq:d2}
 8T^2X(x) + Y(x)\widehat{q}(x) + U(x) q(x) + Z(x)S_1 = 0
\end{equation}
where 
\begin{equation}
\fl
X(x) = J - Kx, \quad
Y(x) = 2C - 4Fx, \quad
U(x) = 4G - 2Cx, \quad
Z(x) = 4D + Cx
\end{equation}
The integral equation (\ref{eq:d2}) can now be transformed into a 
differential equation dividing it by $Y(x)$ and taking the derivative 
with respect to $x$. This leads to the first order differential equation
\begin{equation}
\label{eq:speqd2}
Y(x)\left[U(x) - Y(x)\,x\right] \dot{q}(x) + \mu q(x) + 8T^2\lambda 
+ \nu S_1 = 0 
\end{equation}
with coefficients
\begin{eqnarray}
\lambda &=& \dot{X}Y - X\dot{Y} = -2CK + 4 FJ \\
\mu &=& \dot{U}Y - U\dot{Y} = -4C^2 + 16 FG \\
\nu &=& \dot{Z}Y - Z\dot{Y} =  2C^2 + 16 DF.
\end{eqnarray}
The solution of equation (\ref{eq:speqd2}) reads:
\begin{equation}
\label{eq:qx}
q(x) = \Gamma \frac{x-s}{\sqrt{(x-s)^2 + \Delta}} 
      -a - bS_1, \qquad
0\leq x\leq x_c,
\end{equation}
where
\begin{equation}
a = 8 T^2 \frac{\lambda}{\mu} \qquad
b = \frac{\nu}{\mu} \qquad
s = \frac{C}{2F} \qquad
\Delta = \frac{G}{F} - s^2,
\end{equation}
and we have absorbed a factor $\mu$ into the definition of the integration 
constant $\Gamma$. The quantity $S_1$ is function of $\Gamma$ and $x_c$ 
(and temperature), 
see (\ref{eq:sn}). Introducing the auxiliary function
\begin{eqnarray}
h(z) &=& \int_{0}^{z}\, \rmd x\, \frac{x-s}{\sqrt{(x-s)^2 + \Delta}} 
       + (1-z)\,\frac{z-s}{\sqrt{(z-s)^2 + \Delta}}
\nonumber\\
 &=& \frac{(z-s)(1-s) + \Delta}{\sqrt{(z-s)^2 + \Delta}} - \sqrt{s^2+\Delta}
\end{eqnarray}
this reads
\begin{equation}
S_1 = \frac{\Gamma\, h(x_c) - a}{b -1}
\end{equation}

The value of $\Gamma$, and $x_c$, is determined from equations 
(\ref{eq:speqd1}) and (\ref{eq:d2}). Replacing in equation (\ref{eq:d2}) 
$q(x)$ with the expression (\ref{eq:qx}) yields a
linear equation for $\Gamma$. This can be readily solved noticing 
that since $\Gamma$ does not depend on $x$ we can just set $x=0$ and use 
the identity $\widehat{q}(0) = S_1$. This leads to
\begin{equation}
\label{eq:gamma}
\Gamma = \frac{\Gamma_0}{\Gamma_1 + \Gamma_2\,h(x_c)}
\end{equation}
where
\begin{eqnarray}
\Gamma_0 &= 4 T^2 J (b-1) + a\,(2 G - C - 2D), \nonumber\\
\Gamma_1 &= 2 G (b-1)\frac{s}{\sqrt{s^2+\Delta}}, \\
\Gamma_2 &= 2 G b - C - 2D. \nonumber
\end{eqnarray}

Finally the value of $x_c$, for a given the temperature $T$, is determined
from (\ref{eq:speqd1}). Again we can take advantage of the fact that 
$x_c$ does not depend to $x$ and choose in (\ref{eq:speqd1}) a suitable 
value for $x$, e.g., $x=x_c$ or $x=0$. 
Setting $x=0$ into (\ref{eq:speqd1}) a straightforward algebra 
leads to the equation
\begin{eqnarray}
\fl
2 N + \frac{1}{6 T^2}\bigl[3 J q(0) + (R + 3 K)S_1\bigr]
\nonumber\\
 + \frac{1}{48 T^4}\bigl[ 6 G q(0)^2 + (6C + 12 D) S_1 q(0) + 
\nonumber\\
                       (B - 3C - 3D - 3E + 6F) S_1^2 
                     + (C - 2 H) S_2
                \bigr] = 0
\label{eq:xc}
\end{eqnarray}
where
\begin{eqnarray}
\fl
S_2 = -\Gamma^2\left[-(1-x_c)\frac{\Delta}{(x_c - s)^2+\Delta}
                       + I_2(x_c) - I_2(0) + 1 - 
                       \frac{b(b-2)}{(b-1)^2}h(x_c)^2\right]
\nonumber\\
+ 2 \frac{a}{(b-1)^2} \Gamma h(x_c) - \left(\frac{a}{b-1}\right)^2
\end{eqnarray}
and 
\begin{equation}
\fl
I_2(x) = -\int \,\rmd x\,\frac{\Delta}{(x-s)^2+\Delta}
       = -\sqrt{\Delta}\, \tan^{-1}\left(\frac{x-s}{\sqrt{\Delta}}\right), 
       \qquad \Delta > 0
\end{equation}

\begin{figure}
\includegraphics[scale=1.0]{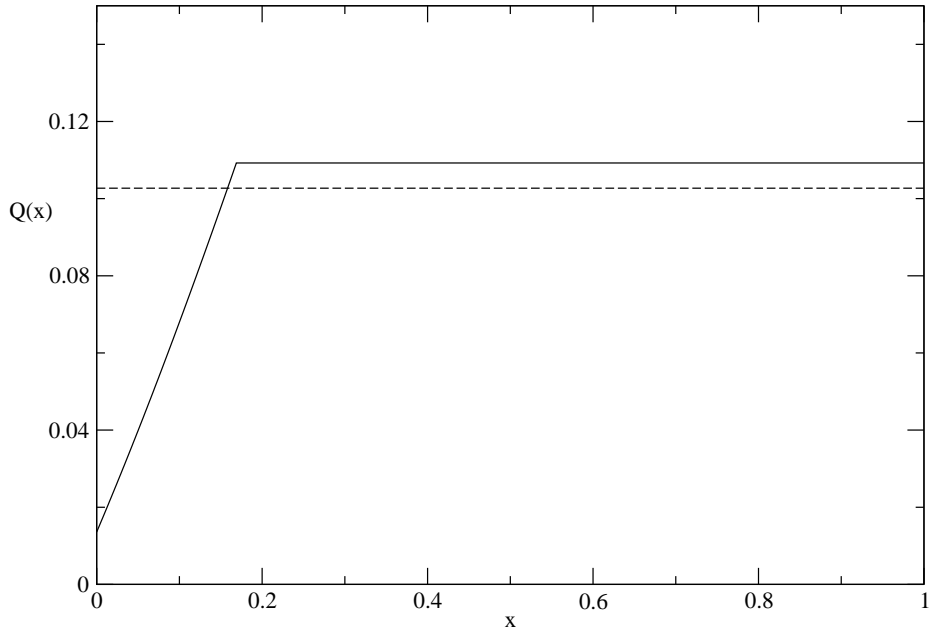}
\caption{$Q(x)$ versus $x$ at temperature $T=0.9$. The horizontal dashed line
         shows the SK solution $Q(x) = q$. For this 
         temperature we have $x_c = 0.168846\ldots$, $Q(x_c) = 0.109238\ldots$ 
         $Q(0) = 0.013570\ldots$ and $q = 0.102701\ldots$. 
         }
\label{fig:q09}
\end{figure}

\begin{figure}
\includegraphics[scale=1.0]{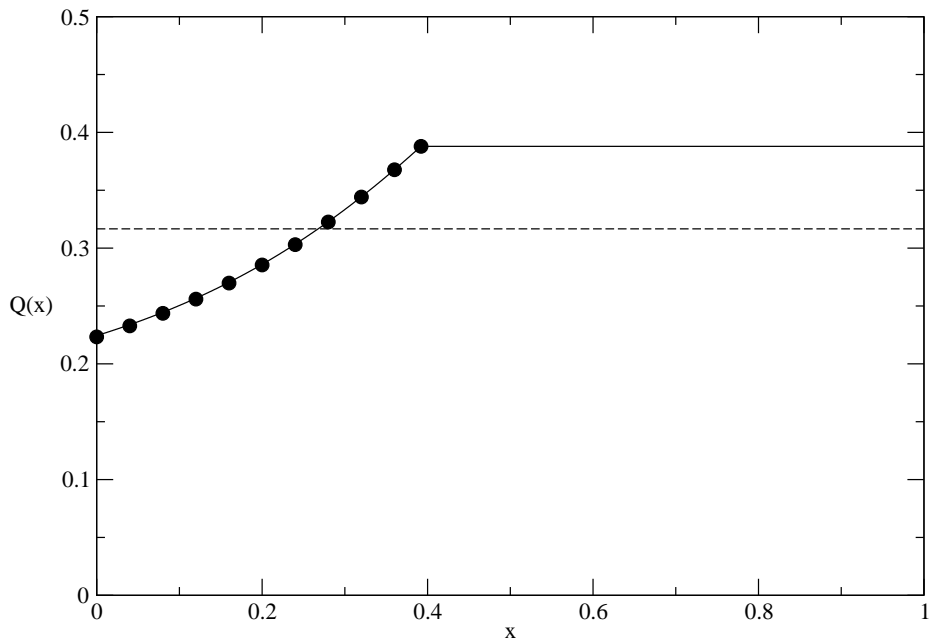}
\caption{$Q(x)$ versus $x$ at temperature $T=0.7$. 
         The full line is the result from the expansion around $T=1$
         to order $\Or(\tau^{13})$, while
         the circle are obtained from the numerical solution of 
         equation (\protect\ref{eq:speqd1}). The horizontal dashed line
         shows the SK solution $Q(x) = q$. 
         For this temperature we have $x_c = 0.3920(6)$, 
         $Q(x_c) = 0.3879(1)\ldots$ 
         $Q(0) = 0.2232(5)\ldots$ and $q = 0.3166(5)\ldots$. 
         }
\label{fig:q07}
\end{figure}

Solving equation (\ref{eq:xc}) for $x_c$ at fixed $T$ yields the value of
$x_c(T)$, that substituted back gives the solution
$q(x)$ as function of temperature. In figures \ref{fig:q09} and
\ref{fig:q07} we show the solutions $Q(x) = q + q(x)$ 
for two different temperatures.

\begin{figure}
\includegraphics[scale=1.0]{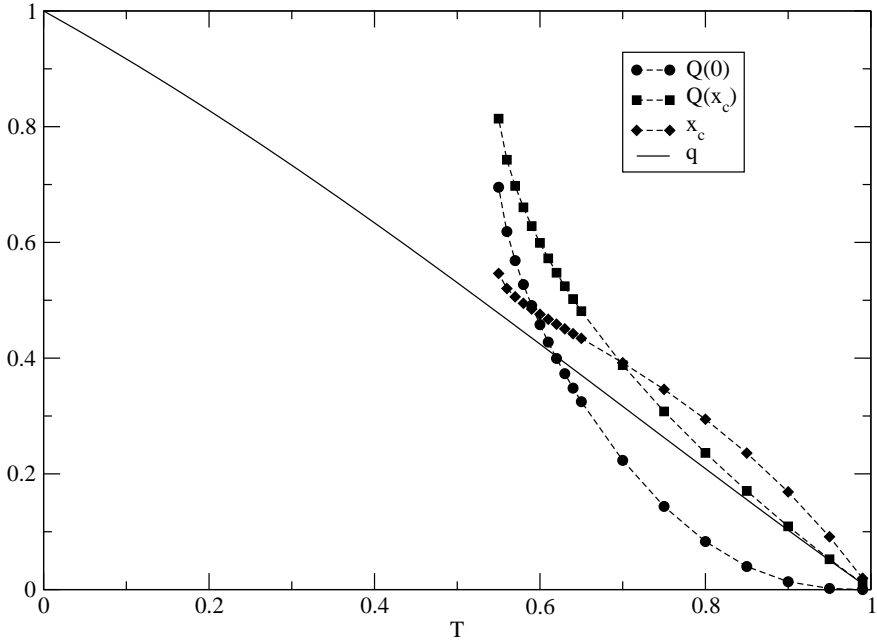}
\caption{$Q(0)$, $Q_(x_c)$ and $x_c$  as function of temperature.
         The full line is the SK result $q=\overline{\theta^2}$.
         The Replica Symmetry broken solution ends at temperature 
         $T=0.549\ldots$.
         }
\label{fig:q0-vs-T}
\end{figure}

From figures one clearly sees that $Q(x=0) \not= 0$.
It grows as the temperature decreases, and overcomes $q$ for
$T < 0.618\ldots$, see also figure \ref{fig:q0-vs-T}.
Retaining in the expansion of $\Omega[Q]$ only terms up to order 
$\Or(q_{ab}^4)$ 
breaks the replica symmetry, however,
this approximation is not good enough to change the SK result
$Q(x=0) = q\not=0$ to the expected one $Q(x=0)=0$.\footnote{$Q(x=0)$ 
must vanish in absence of external fields that break the up/down
symmetry.  For instance in the $q\geq 4$ Potts model the symmetry is 
broken and indeed $Q(0)\not=0$}
To recover the latter one has to add more terms in the expansion, 
probably all terms.

Below temperature $T=0.549\ldots$ equation (\ref{eq:speqd1}) ceases to have a 
physical solution and only the SK solution $Q(x) = q$
survives. In figure \ref{fig:q0-vs-T} we show the values of $Q(0)$, $Q(x_c)$ and
$x_c$ as function of temperature.

\subsection{Solution near $T_c=1$}
Near the critical temperature $T_c = 1$, where both $q$ and $q_{ab}$ vanish,
the solution of equation (\ref{eq:speqd1}) can be found as a series expansion 
in the (small) parameter $\tau = T_c - T$.
For example to $\Or(\tau^5)$ we have
\begin{equation}
  x_c = 2\tau - 4\tau^2 + \frac{40}{3}\tau^3 - \frac{665}{9}\tau^4
        + \frac{68567}{135}\tau^5 + \Or(\tau^6)
\end{equation}

\begin{equation}
Q(0) = q + q(0) =  \frac{56}{3}\tau^3  
                 - \frac{220}{3}\tau^4 
                 + \frac{3968}{9}\tau^5
                 + \Or(\tau^6) 
\end{equation}

\begin{equation}
Q(x_c) = q + q(x_c) =   \tau
                      + \tau^2 
                      - \tau^3 
                      + \frac{5}{2}\tau^4
                      - \frac{413}{90}\tau^5
                      + \Or(\tau^6) 
\end{equation}

The resulting series are not convergent, but can be handled by using the Pad\'e
Approximants. We note that the series expansion of $x_c$ has the form of
a Stieltjes series $\sum a_n(-\tau)^n$. For these series it is known that 
the diagonal Pad\'e approximant $P_N^N(\tau)$ gives an upper bound and 
the approximant $P_{N+1}^N(\tau)$ a lower bound of the sum
\cite{BenOrs99}. 
Moreover in the limit of large $N$ both approximants converge, and if
they converge to the same limit this is the value of the sum.
By using the Pad\'e approximants we were able to use the series expansion 
almost everywhere in the low temperature phase, where the Replica Symmetry 
broken solution exists. 
For example for temperature $T=0.7$ by using the series expansion to 
$\Or(\tau^{13})$ we have  $x_c = 0.3920(6)$, the error being estimated from the
difference between the Pad\'e Approximants $P_N^N$ and $P_{N+1}^N$.
A comparison between the numerical and the power series solutions is shown
in figure \ref{fig:q07}, the agreement is rather good.

\section{Expansion around the Paramagnetic Solution:}
\label{sec:expa0}
At the critical point $T_c = 1$ the order parameter function $Q(x)$ vanishes, 
one can then think of expanding the functional $\Omega[Q]$ around $Q_{ab}=0$, 
i.e., the paramagnetic solution. 
Such an expansion, first considered by Bray and Moore \cite{BraMoo79},
is at the basis of the so-called {\sl Truncated Model} \cite{Parisi79}
largely used to study the properties of the solution $Q(x)$ near the critical 
point. 
See \cite{SchTatCht06} for an extension to more general models.
Despite it usefulness, 
the Truncated Model is a poor approximation for the SK model.
Indeed, in the same spirit of the Landau Theory of second order
transition, it retains only the main mathematical structure of the 
order $\Or(Q_{ab}^4)$ expansion of $\Omega[Q]$ around $Q_{ab}=0$,
but with arbitrary coefficients. 
Using the results of Section \ref{sec:expa}, we can investigate the 
properties of the $\Or(Q_{ab}^4)$ approximation of the SK model.

The expansion of the replica free energy functional around $Q_{ab} = 0$ 
is obtained by setting $q=\theta = 0$ in (\ref{eq:freen4c}). This yields
\begin{eqnarray}
\label{eq:freen4c0}
\fl
-n\beta f = n\,\ln 2 
  +\frac{1}{4T^4}
             N\, \sum_{ab}q_{ab}^2
  +\frac{1}{6(2T^2)^3}
  + K\, \sum_{abc} q_{ac}q_{cb}q_{ba}
\nonumber\\
  +\frac{1}{24(2T^2)^4}
            \left[
   F\,\sum_{abcd}q_{ab}q_{bc}q_{cd}q_{da}
   + G\,\sum_{ab}q_{ab}^4 
   - H\,\sum_{abc}q_{ac}^2q_{cb}^2
   \right]
\end{eqnarray}
with
\begin{equation}
\label{eq:const0}
N = 1 - T^2, \quad K = 8, \quad F = 48, \quad G = 32, \quad H = 96
\end{equation}

Stationarity of (\ref{eq:freen4c0}) with respect to variations of $q_{ab}$ 
leads to the stationary point equation, that for $R\to\infty$ reads:
\begin{eqnarray}
\label{eq:speq0}
\fl
2Nq(x) 
+\frac{K}{4T^2}
        \left(\int_{0}^{x}\rmd y\,\dot{q}(y)\,\widehat{q}(y) + S_1\, q(0)\right)
\nonumber\\
+\frac{1}{24 T^4}
 \Biggl[
   3F\left(\int_{0}^{x}\rmd y\, \dot{q}(y)\,\widehat{q}(y)^2 + S_1^2 q(0)\right)
\nonumber\\
   + G q(x)^3 - HS_2 q(x)
 \Biggr]
= 0.
\end{eqnarray}
Applying the differential operator 
$\widehat{\bm O}= (1/\dot{q}(x))(\rmd/\rmd x)$, 
as done in Section \ref{sec:solu}, reduces the above equation to:
\begin{eqnarray}
\label{eq:speqd10}
\phantom{0} &\phantom{=}&
2N
+ \frac{K}{2T^2}\,\widehat{q}(x)
+ \frac{1}{24T^4}\Bigl[
        3F\widehat{q}(x)^2 + 3Gq(x)^2 - HS_2\Bigr]
= 0.
\end{eqnarray}
This equation is not yet simple enough to be solved. A second application of
$\widehat{O}$ leads to
\begin{equation}
\label{eq:d20}
-2T^2Kx - F x \widehat{q}(x) + G q(x) = 0.
\end{equation}
Dividing this equation by $F x$ and taking the derivative with respect to
$x$ transform the integral equation (\ref{eq:d20}) into a differential 
equation, that solved yields:
\begin{equation}
\label{eq:qx0}
  q(x) = \Gamma \frac{x}{\sqrt{x^2 + G/F}}, \qquad 0\leq x\leq x_c
\end{equation}
with
\begin{equation}
 \Gamma = 2 T^2\frac{K}{F}\frac{\sqrt{x_c^2 + G/F}}{x_c + G/F}
\end{equation}
determined inserting the form (\ref{eq:qx0}) 
of $q(x)$ into (\ref{eq:d20}).

The end point $x_c$ is not a free parameter and must be determined as function
of temperature from (\ref{eq:speqd10}). 
Inserting $q(x)$ from (\ref{eq:qx0}) and 
\begin{equation}
\widehat{q}(x) = \frac{G}{F} \frac{q(x)}{x} - 2 T^2 \frac{K}{F}
\end{equation}
into (\ref{eq:speqd10}) one ends up with the following equation
\begin{eqnarray}
\label{eq:xc0}
\fl
2N - \frac{K^2}{2F} + \frac{K^2}{6F}\frac{x_c^2 + G/F}{(x_c+G/F)^2}
\Biggl[ 3 G + 
\nonumber\\
  H\left(
          1 
        - \sqrt{\frac{G}{F}}\, \tan^{-1}\left(\sqrt{\frac{F}{G}}x_c\right)
        - \frac{G}{F}\frac{1-x_c}{x_c^2 + G/F}
              \right)
\Biggr] = 0
\end{eqnarray}
that solved for $x_c$ gives the value of $x_c(T)$. 

At the critical temperature $T=T_c=1$, where $N=0$, 
$x_c$ vanishes and increases as the temperature is decreased below $T_c$. 
Introducing the small parameter $\tau = 1 -T$, the solution of equation 
(\ref{eq:xc0}) can be expressed as a power series. For example to 
order $\Or(\tau^5)$ we have,
\begin{equation}
 x_c =   2 \tau 
       + 12 \tau^2
       + \frac{280}{3} \tau^3
       + \frac{2437}{3} \tau^4
       + \frac{37641}{5} \tau^5
       + \Or(\tau^6)
\end{equation}
\begin{equation}
Q(x_c) =   \tau 
         + \tau^2
         + \frac{44}{3} \tau^3
         + \frac{701}{6} \tau^4
       + \frac{30763}{30} \tau^5
       + \Or(\tau^6)
\end{equation}
and
\begin{equation}
\Gamma =   \frac{1}{\sqrt{6}}
         - \frac{5}{\sqrt{6}} \tau
         + \frac{1}{\sqrt{6}} \tau^2
         - \frac{23}{\sqrt{6}} \tau^3
         - 51\sqrt{\frac{3}{2}} \tau^4
         - \frac{2972}{5}\sqrt{\frac{2}{3}} \tau^5
         + \Or(\tau^6).
\end{equation}
Notice that in this case $Q(x=0)=0$.

The maximum allowed value of $x_c$
is $1$. Setting $x_c=1$ into (\ref{eq:xc0}), and replacing the constants
by their values (\ref{eq:const0}), we find that 
the Replica Symmetry Broken solution (\ref{eq:qx0}) becomes non-physical 
below the temperature
\begin{equation}
 T_{\rm FRSB} = \frac{1}{3}\sqrt{\frac{2}{5}
          \left[21 - \sqrt{6}\,
          \tan^{-1}\left(\sqrt{\frac{3}{2}}\right)
                      \right]}
   = 0.9148\ldots
\end{equation}
where $x_c>1$. At this temperature $Q(x_c)$ reaches its maximum value
\begin{equation}
\lim_{T\to T_{\rm FRSB}} Q(x_c) = \frac{2}{225}\left[
            21 - \sqrt{6}\,\tan^{-1}\left(\sqrt{\frac{3}{2}}\right)
            \right] = 0.16737\ldots
\end{equation}
We conclude this Section noticing that 
for temperatures above. but close to, $T_{\rm FRSB}$ equation (\ref{eq:xc0}) 
can be solved as power series of $1-x_c$. We do not report the expansion here.

\section{Discussion and Conclusions:}
\label{sec:conc}
In this work, we have derived the expansion of the Sherrington-Kirkpatrick 
model replica free energy functional around the Replica Symmetric (RS) solution
$Q^{({\rm RS})}_{ab} = \delta_{ab} + q(1-\delta_{ab})$.
We have considered in detail the approximation obtained by truncating the
expansion to fourth order in $Q_{ab}-Q^{({\rm RS})}_{ab}$, i.e., the lowest 
nontrivial approximation to have a continuous Replica Symmetry Breaking.
The stationarity equation (\ref{eq:speq}) associated with the approximate 
free energy functional (\ref{eq:freen4c})  can be solved and the explicit 
form of the 
Full Replica Symmetry Broken (FRSB) solution $Q(x)$, for $0\leq x\leq x_c$, 
can be determined. The FRSB solution appears at the critical 
temperature $T_c=1$, as the RS solution, and exists only down to the 
finite temperature $T=0.549\ldots$. Below only the RS solution survives.

A peculiar feature of the FRSB solution is that 
$Q(x=0)\not=0$, and vanishes as $(T_c - T)^3$ as the temperature $T$ 
approaches the critical temperature $T_c=1$.
This property can be traced back to the fact that the FRSB solution
``opens'' around the RS solution $Q^{({\rm RS})}(x)$, i.e.,
$Q(x=0) < q < Q(x_c)$, as the temperature decreases below the critical
temperature $T_c$. As the temperature is decreased below $T_c$ the RS solution 
$q$ increases and drags $Q(x=0)$ to finite values. We note that 
at $T=0.618\ldots$ 
the value of $Q(x=0)$ eventually overcomes that of $q$.

Setting $q=0$ one recovers the expansion of the replica free energy 
functional around the paramagnetic solution $Q^{({\rm PM})}_{ab} = \delta_{ab}$
to order $\Or(Q_{ab}^4)$.
This turns out to be rather interesting because such an expansion is 
at the basis of the Truncated Model used to study the properties of the
FRSB solution $Q(x)$ near the transition. 
To our knowledge, a study of this approximation with the correct 
coefficients of the expansion was never done.
Indeed the Truncated Model, and the one in which one keeps all terms
generated by the expansion of $\Omega[Q]$ to order $\Or(Q_{ab}^4)$, have been
studied with arbitrary coefficient. As consequence of this
the existence od a FRSB solution was always taken for granted, but 
never verified. We have studied the existence of the FRSB solution
for this expansion in the last part of this work. Surprisingly 
it turns out that the FRSB solution exists only close to the critical 
temperature, in the range of temperature $0.9148\ldots\leq T\leq 1$.
Therefore such expansions truncated to the forth order 
cannot be used to study the solution of the SK model near zero temperature.

To summarize,
\begin{itemize}
\item The expansion (to $4^{\rm th}$ order) around the (Replica Symmetric) SK 
solution $Q_{ab}^{({\rm RS})} = \delta_{ab} + q(1-\delta_{ab})$ yields, in the
limit $R\to\infty$, a $Q(x)$ that does not vanish for $x$ null, in contrast 
with the exact Parisi Solution. The solution exists only in the range
$0.549\ldots\le T\le T_c=1$.

\item The expansion (to $4^{\rm th}$ order) around the paramagnetic solution
$q=0$, yields a $Q(x)$ that does vanish for $x$ null. But it exists only 
close to $T_c = 1$, $0.915\ldots\le T\le T_c=1$.
\end{itemize}

In this work we have studied the existence of FRSB solutions.
Finite RSB solutions may also exist. 
These, however, may exhibit problems similar to 
those found for the FRSB solution. 
For example, inserting the Replica Symmetric 
{\it Ansatz} $q_{ab} = q\, (1-\delta_{ab})$ into the free energy functional 
(\ref{eq:freen4c0}), and taking the limit $n\to 0$, or expanding the SK free 
energy  (\ref{eq:freenSK}) around $q=0$ to the fourth order in $q$, one ends
up with
\begin{equation}
 -\beta f = \ln 2 - \frac{1-T^2}{4 T^4}\,q^2
                  + \frac{1}{3 T^6}\, q^3
                  -\frac{17}{24 T^8}\, q^4.
\end{equation}
Stationarity respect to variations of $q$ yields 
the paramagnetic solution $q=0$ and the RS solution
\begin{equation}
 q = \frac{T^2}{17}\left[3 - \sqrt{3(17 T^2 - 14)}\right].
\end{equation}
The latter correctly vanishes at the critical point $T=T_c=1$, 
but exists only down to temperature  $T=\sqrt{14/17}\simeq 0.907\ldots$,
slightly below the lower end $T=0.915\ldots$ of the FRSB solutions,  
where the quantity under the square root becomes negative. 

The situation is only slightly better considering the expansion around the 
Replica Symmetric  SK solution since now the RS solution exists down to $T=0$.
The RS {\it Ansatz} 
$q_{ab} = \delta q\,(1 -\delta_{ab})$ yields indeed, besides the trivial 
solution $\delta q=0$, a $\delta q\not= 0$ solution leading to the 
unphysical result
$Q=q + \delta q \simeq -\frac{31}{3}(1-T)^3$ as $T\sim 1^-$.
The reason is that the expansion around the SK solution includes the 
contribution of more diagrams: all diagrams that are needed to build the SK 
free energy. In this sense
this expansion is a better approximation for the SK model, 
as also reflected by the larger temperature range were the FRSB solution exist.

The improvement is only apparent since for both expansions the 
RS solution, as well as the Paramagnetic solution for $T<1$, 
has a negative Replicon mass and is, hence, unstable. 
For what concerns the FRSB solution, whether in the full expansion 
(\ref{eq:freen4c0}) or in the truncated model, they both have the same 
stability properties in the most dangerous sector, i.e., in the 
Replicon subspace (by virtue of the Ward-Takahashi identities
\cite{DeDomTemKon98}).
Thus the FRSB, where it does exist,
is marginally stable with null Replicon masses. We believe that 
this feature remains true for the expansion around the SK solution as well.

Despite this, the limited range of temperature (and not including $T=0$)
in which these expansions to $4^{\rm th}$ order exist, makes them of little
help  to study the properties of 
the SK model near zero temperature.
To extend the range of validity one should retain more terms in 
the expansion, and probably all terms (or infinite subseries thereof)
since the particular structures of the 
expansion 
(in powers of $\beta$) may otherwise 
lead to difficulties for very low temperatures.
We observe that to overcome this problem a construction based upon an 
expansion around a spherical 
approximation, which leads instead to an expansion in $T$,
has been recently proposed \cite{CriDeDomSar09}.

\ack
A.C. would like to thank the IPhT of CEA, 
where part of this work was done, for the kind hospitality and support.

\appendix

\section{The Truncated Model}
The Truncated model is defined by the free energy
\begin{equation}
\label{eq:freentr}
-n f =
   \frac{\tau}{2} \sum_{ab}q_{ab}^2
  + \frac{w}{6} \sum_{abc} q_{ac}q_{cb}q_{ba}
  +\frac{u}{12} \sum_{ab}q_{ab}^4 
\end{equation}
with $\tau = 1 - T$ and $w$ and $u$ arbitrary and positive.
Comparison of (\ref{eq:freentr}) and (\ref{eq:freen4c0}) shows that
\begin{equation}
\label{eq:consttr}
\tau = \frac{N}{2T^3}, \quad w = \frac{K}{8T^5}, \quad u = \frac{G}{32T^7} 
\end{equation}
while $F=H=0$.
We can then read the equation for $q(x)$ directly from Section \ref{sec:expa0}.
Setting $F=0$ in (\ref{eq:d20}) one readily obtains the known linear form
of $q(x)$ for the truncated model:
\begin{equation}
 q(x) = 2T^2\frac{K}{G}x 
      = \frac{w}{2u} x, \qquad 0\leq x_c\leq x_c
\end{equation}
Finally setting $F=H=0$ in (\ref{eq:speqd10}), and using the above 
linear form of $q(x)$, yields
\begin{equation}
\label{eq:xctr}
2 N - \frac{K^2}{G} x_c + \frac{K^2}{2G} x_c^2 = 0
\end{equation}
that gives $x_c$ as function of temperature. 
The value of $x_c$ is zero for $T=1$ and increases as $T$ decreases below $1$.
By setting $x_c=1$ into (\ref{eq:xctr}) leads to the critical temperature
\begin{equation}
  T_{\rm trm} = \sqrt{1 - \frac{K^2}{4G}}
\end{equation}
below which the Replica Symmetry broken solution ceases to exist. If 
$1-K^2/4G < 0$ the solution exists down to $T=0$. If we use the values
$K=8$ and $G=32$ we have $T_{\rm trm} = 1/\sqrt{2} = 0.707\ldots$.

We note that due to the presence of $T$-factors in the relation 
between $(N,K,G)$ and $(\tau,w,u)$ the critical temperature has a slightly 
different form if expressed in the latter:
\begin{equation}
  T_{\rm trm} = 1 - \frac{w^2}{4u}
\end{equation}
and is valid if $w$ and $u$ are temperature independent.

\section{Terms $\Or(q_{ab}^2)$: details}
\label{app:O2det}
The terms of order $\Or(q_{ab}^2)$ are given by
\begin{equation}
\label{eqd:o2}
\left\langle\left(\sum_{ab}q_{ab}\,\sigma_a\sigma_b\right)^2\right\rangle = 
    \sum_{ab\atop cd} q_{ab}\,q_{cd}\,
          \langle\sigma_a\sigma_b\sigma_c\sigma_d\rangle
\end{equation}
To evaluate this term we have to find all possible ways
of equating the $ab$ indexes to $cd$ indexes, with the constraint
$a\not= b$ and $c\not= d$ since $q_{aa}=0$. In the following to denote that 
a group of indexes must be all different we shall write them in parenthesis, 
hence in the present case we have two write $(ab)$ and $(cd)$.

We clearly have three 
possible cases: all different, one equal, two equals. For later use
it is useful to represent them graphically. 
If we denote $q_{ab}$ by a straight line then the case of
all different indexes is represented as
\begin{equation}
\includegraphics[bb= 106 656 140 713]{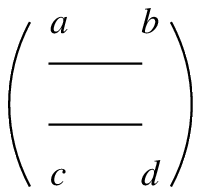}
\end{equation}
and the value of the average is $\overline{\theta^4}$ since all spin indexes 
are different.
Next there are four possible ways of equating one $(ab)$ index to one $(cd)$ 
index. These are 
\begin{equation}
\label{eqd:o21e}
\fl
\includegraphics[bb= 106 656 324 713]{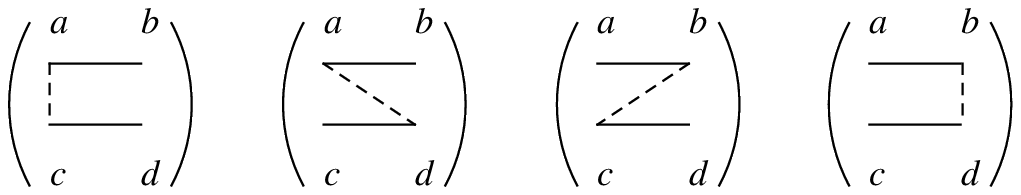}
\end{equation}
where the indexes connected by a dashed line are equal. In this case two 
indexes in the average are equal, so only two spins survive and the spin 
average gives $\overline{\theta^2}$. 

Finally there are two possible way of equating indexes $(ab)$ and indexes
$(cd)$ with the constraint $a\not=b$ and $c\not= d$ and reads
\begin{equation}
\label{eqd:o22e}
\includegraphics[bb= 106 656 324 713]{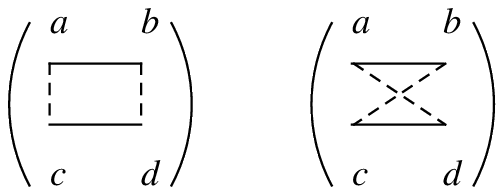}
\end{equation}
In this cases we have two pairs of equal indexes in the average, so all spins 
disappears and the average give $1$.

To evaluate (\ref{eqd:o2}) we have to sum each diagram over $a,b,c,d$, then 
it is easy to realize that since the matrix $q_{ab}$ is symmetric all
four diagrams in (\ref{eqd:o21e}) give the same contribution, and so do the
two diagrams in (\ref{eqd:o22e}). These will be denoted as
\begin{equation}
\label{eqd:o22ea}
\includegraphics[bb= 106 656 324 713]{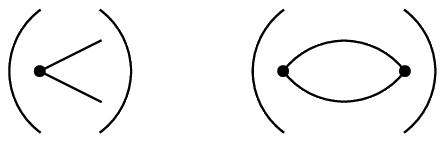}
\end{equation}
respectively.

Collecting all terms we have
\begin{equation}
\fl
\label{eqd:q2rest}
\left\langle\left(\sum_{ab}q_{ab}\,\sigma_a\sigma_b\right)^2\right\rangle =
      \overline{\theta^4}\,\sum_{(abcd)} q_{ab}\,q_{cd}
  + 4\,\overline{\theta^2}\,\sum_{(abc)} q_{ac}\,q_{cb}
  + 2 \sum_{(ab)} q_{ab}^2
\end{equation}
The restricted sums can be transformed into unrestricted sums by inserting a 
factor $(1-\delta_{ab})$ for each pair of indexes $(ab)$ to enforce the 
constraint and removing the constraint over the indexes. 
By expanding now the resulting products of $(1-\delta)$'s each restricted sum 
is finally expressed as a combination of unrestricted sums. 
Diagrammatically we have
\begin{equation}
\label{eqd:o2adru}
\includegraphics[bb= 97 658 311 695]{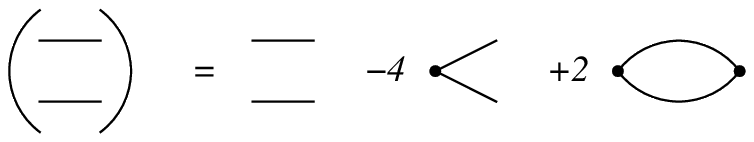}
\end{equation}
and
\begin{equation}
\label{eqd:o21eur}
\includegraphics[bb= 97 658 250 695]{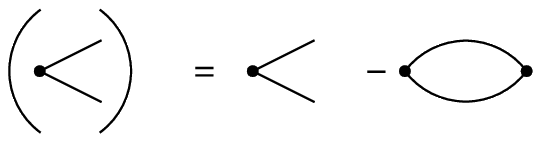}
\end{equation}
Inserting these expressions into (\ref{eqd:q2rest}) after simple 
manipulations we end up with
\begin{equation}
\fl
\label{eqd:q2unrest}
\left\langle\left(\sum_{ab}q_{ab}\,\sigma_a\sigma_b\right)^2\right\rangle =
  \overline{\theta^4}\ \includegraphics[bb= 106 674 125 695]{q2adu}
+ 4\,\overline{\theta^2(1-\theta^2)}\ %
\includegraphics[bb= 105 675 126 694]{q21eu}
  + 2\,\overline{(1-\theta^2)^2}\ %
\includegraphics[bb= 106 683 147 702]{q22eu}
\end{equation}
The first term is disconnected and hence it does not contribute to the free 
energy, therefore to order $\Or(q_{ab}^2)$ the free energy reads
\begin{eqnarray}
\fl
\label{eqd:freenc}
-n\beta f = -n\beta f_{\rm SK} 
  -\frac{\beta^2}{4}
            \left[1-\beta^2\overline{(1-\theta^2)^2}\right]\sum_{ab}q_{ab}^2 
\nonumber\\
 + \frac{\beta^4}{2}\overline{\theta^2(1-\theta^2)}
         \sum_{abc}q_{ac}q_{cb} 
+ \Or(q_{ab}^3)
 \end{eqnarray}


\begin{thebibliography}{99}

\bibitem{SheKir78}
  D. Sherrington and S. Kirkpatrick 1978
  {\it Phys. Rev.} B {\bf 17} 4384

\bibitem{Parisi79}
  G. Parisi 1979
  {\it Phys. Rev. Lett.} {\bf 43} 1754

\bibitem{Parisi80}
  G. Parisi 1980
  {\it J. Phys.} A {\bf 13} 1101

\bibitem{Duplantier81}
  B. Duplantier 1981
  {\it J. Phys.} A. {\bf 14} 283

\bibitem{Parisi83}
  G. Parisi 1983
  {\it Phys. Rev. Lett.} {\bf 50} 1946

\bibitem{CriRiz02}
  A. Crisanti and T. Rizzo 2002
  {\it Phys. Rev.} E {\bf 65} 46137

\bibitem{CriRizTem03}
  A. Crisanti, T. Rizzo and T. Temesvari,
  Eur. Phys. J. B {\bf 33}, 203 (2003)

\bibitem{Pankov06}
  S. Pankov 2006
  {\it Phys. Rev. Lett.} {\bf 96} 197204

\bibitem{OppShe05}
  R. Oppermann and D. Sherrington 2005
  {\it Phys. Rev. Lett.} {\bf 95} 197203

\bibitem{OppSchShe07}
  R. Oppermann, M.J. Schmidt and D. Sherrington 2007
  {\it Phys. Rev. Lett.} {\bf 98} 127201

\bibitem{SchOpp08}
  M.J. Schmidt and R. Oppermann 2008
  {\it Phys. Rev.} E {\bf 77} 061104

\bibitem{BraMoo78}
  A.J. Bray and M.A. Moore 1978
  {\it Phys. Rev. Lett.} {\bf 41} 1068

\bibitem{DeDomKon83}
  C. De Dominicis and I. Kondor 1983
  {\it Phys. Rev.} B {\bf 27} 606

\bibitem{BraMoo79}
  A. Bray and M. Moore 1979
  {\it J. Phys.} C {\bf 12} 79

\bibitem{deAlmTho78}
  J. R. de Almeida and D. J. Thouless 1978
  {\it J. Phys.} A {\bf 11} 983

\bibitem{DeDomTemKon98}
  C. De Dominicis, T. Temesvari and I. Kondor 1998
  {\it J. de Physique IV France} {\bf 8} 13  
  (Preprint {\it cont-mat}/9802166)
  Equation numbering has been messed up at the editing stage,
  the reader should rather consult the cond-mat version.

\bibitem{DeDomCarTem97}
  C. De Dominicis, D.M. Carlucci and T. Temesvari 1997
  {\it J. Phys. I France} {\bf 7} 105

\bibitem{MezPar91}
  M. Mezard and G. Parisi 1991
  {\it J. Phys. I France} {\bf 1}, 809

\bibitem{SchTatCht06}
  T.I. Schelkacheva, E.E. Tateyeva and N.M. Chtchelkatchev 2006
  {\it Phys. Lett. A} {\bf 358}, 222

\bibitem{BenOrs99}
  C. Bender and S.A. Orszag
  {\it Advanced Mathematical Methods For Scientists and Engineers}
  (Springer, 1999)

\bibitem{CriDeDomSar09}
  A. Crisanti, C. De Dominicis and T. Sarlat 2009
  {\it Eur. Phys. J. } B (in press); 
  {\it cond-mat}/0909.2556

\end{thebibliography}
\end{document}